\documentclass[twocolumn,aps,prb,a4]{revtex4-1}
\usepackage{amsmath}
\usepackage{amssymb}
\usepackage{braket}
\usepackage{graphicx}
\usepackage[table]{xcolor}
\usepackage{verbatim}
\usepackage{float}
\usepackage{bbold}
\usepackage{multirow}
\usepackage{color}
\begin{document}
\title{Spin-orbit coupling, optical transitions, and spin pumping in mono- and few-layer InSe}
\author{S. J. Magorrian}
\author{V. Z\'{o}lyomi}
\author{V. I. Fal'ko}
\affiliation{National Graphene Institute, University of Manchester, Booth St E, Manchester, M13 9PL, United Kingdom}
 \begin{abstract}
We show that spin-orbit coupling (SOC) in InSe enables the optical transition across the principal band gap to couple with in-plane polarized light. This transition, enabled by $p_{x,y}\leftrightarrow p_z$ hybridization due to intra-atomic SOC in both In and Se, can be viewed as a transition between two dominantly $s$- and $p_z$-orbital based bands, accompanied by an electron spin-flip. Having parametrized $\mathbf{k\cdot p}$ theory using first principles density functional theory we estimate the absorption for $\sigma^{\pm}$ circularly polarized photons in the monolayer as $\sim 1.5\%$, which saturates to $\sim 0.3\%$ in thicker films ($3-5$ layers). Circularly polarized light can be used to selectively excite electrons into spin-polarized states in the conduction band, which permits optical pumping of the spin polarization of In nuclei through the hyperfine interaction. 
\end{abstract}
\maketitle
\section{Introduction}
Two-dimensional (2D) materials, such as atomic layers of transition metal dichalcogenides \cite{Mak2010,Splendiani2010,Korn2011,Wang2012,Xu2014,Jones2013,Gan2013,Wu2014,Sie2015,Wang2015,Liu2015} and metal chalcogenides GaSe \cite{zhou2015strong,jie2015layer,hu2012synthesis,late2012gas,lei2013synthesis,aziza2017tunable,Pozo2015,jung2015}, GaTe \cite{liu2014gate,hu2014highly,huang2016plane}, and InSe \cite{Lei2014,Mudd2013,Mudd2014,Mudd2015,Tamalampudi2014,Balakrishnan2014,yan2017fast} are attracting a lot of attention due to their promise for applications in optoelectronics. This is based on observations of their optical properties, which can differ strongly from those of their parent bulk materials, and which have demonstrated room-temperature electroluminesence\cite{Balakrishnan2014}, strong photoresponsivity\cite{Lei2014} with a broad spectral response\cite{Mudd2015,Tamalampudi2014,yan2017fast}, and band gap tunability\cite{Mudd2013}. Recent studies of luminescence \cite{bandurin2017high} and magnetoluminescence \cite{SciRep_2016} in InSe have shown a strong dependence of the band gap on the number of layers, from $\sim2.8$~eV for the monolayer to $\sim1.3$~eV for thick films. These experiments have identified  two main photoluminescence lines, interpreted\cite{bandurin2017high} as a lower energy transition between bands dominated by $s$ and $p_z$ orbitals (A-line) and hot luminescence, involving holes in a deeper valence band based on $p_x$ and $p_y$ orbitals (B-line). The band structure analysis of mono- and few-layer InSe \cite{zhuang2013single,rybkovskiy2014transition,magorrian2016electronic,*tb_erratum,Zolyomi2014,zhou2017multiband,debbichi2015two,jalilian2017electronic,xiao2017defects,demirci2017structural,wick2015electronic,cai2017charge,hung2017thermo,Do2015,Li2015,Nissim2017,Ayadi2017,Li2017} has revealed that the conduction and valence band edges near the $\Gamma$-point are non-degenerate, being dominated by $s$ and $p_z$ orbitals of both metal and chalcogen atoms. Combined with the opposite $z\rightarrow -z$ (mirror reflection) symmetry of conduction and valence bands, this determines that the transition across the principal band gap has a dominantly electric dipole-like character, coupled to out-of-plane polarized photons. In contrast, the B-line is found to be related to the recombination of hot holes in a twice-degenerate valence band based on $p_x$ and $p_y$ orbitals, and this transition is strongly coupled to in-plane polarized light. These selection rules are important in understanding experimental observations of these transitions, in particular where the experiments are carried out at normal incidence or emission, since the polarization of the incident/emitted light is then necessarily parallel to the plane of the 2D crystal.

Spin-orbit coupling (SOC) in indium and selenium is capable of igniting additional transitions, accompanied by an electron spin-flip, with polarization properties and selection rules which differ from the selection rules for transitions in the absence of SOC, and which are determined by the angular momentum transfer from the photon to the spin of the $e$-$h$ excitation. Here, we use $\mathbf{k\cdot p}$ theory to show how SOC ignites spin-flip transitions between the conduction and valence band edges in monolayer InSe, coupled to in-plane polarized light, and we evaluate the optical oscillator strength in mono- and few-layer InSe films, which corresponds to an absorption coefficient $\sim 1.5\%$ in the monolayer, and $\sim 0.3\%$ in thicker films ($3-5$ layers).
\section{Monolayer}
To analyse the effect of SOC on the band-edge states and their optical properties in the vicinity of the $\Gamma$-point in monolayer InSe we amend the $\mathbf{k\cdot p}$ Hamiltonian of Ref.~\onlinecite{magorrian2016electronic} by including atomic SOC, $\lambda\mathbf{L\cdot s}$, leading to inter-band ($p_x/p_y\leftrightarrow p_z$) mixing,
\begin{widetext}
\begin{equation}
\hat{H}= \left(
\begin{array}{cccc}
H_c\mathbb{1}_s   & E_zd_z & \frac{e\beta_1}{cm_e}\mathbb{1}_s\otimes\mathbf{A} & \lambda_{c,v_2}\boldsymbol{\hat{s}}\\
E_zd_z & H_v\mathbb{1}_s&\lambda_{v,v_1}\boldsymbol{\hat{s}} & \frac{e\beta_2}{cm_e}\mathbb{1}_s\otimes\mathbf{A} \\
\frac{e\beta_1}{cm_e}\mathbb{1}_s\otimes\mathbf{A^T}& \lambda_{v,v_1}\boldsymbol{\hat{s}}^\mathbf{T} & \mathbb{1}_s\otimes \mathbf{H}_{v_1}+\lambda_{v_1}s_z\otimes\sigma_y  &  0\\
\lambda_{c,v_2}\boldsymbol{\hat{s}}^\mathbf{T}& \frac{e\beta_2}{cm_e}\mathbb{1}_s\otimes\mathbf{A^T}  & 0 &  \mathbb{1}_s\otimes \mathbf{H}_{v_2}+\lambda_{v_2}s_z\otimes\sigma_y  \\

\end{array}
\right);
\label{eq_h0_6x6}
\end{equation}
\begin{equation*}
H_c\approx\hbar^2 k^2/2m_c;\quad H_{v} (k)\approx E_{v} + E_{2}k^{2} + E_{4}k^{4};\quad\mathbf{H}_{v_{1(2)}}\approx\left[E_{v_{1(2)}}+\frac{\hbar^2k^2}{2m_{1(2)}}\right]\mathbb{1}_{\sigma}+\frac{\hbar^2(k_x^2-k_y^2)}{2m_{1(2)}^{\prime}}\sigma_z +\frac{2\hbar^2k_x k_y}{2m_{1(2)}^{\prime}}\sigma_x.
\label{kdotp_v12}
\end{equation*}
\end{widetext}

\begin{figure}
\includegraphics[width=0.48\textwidth]{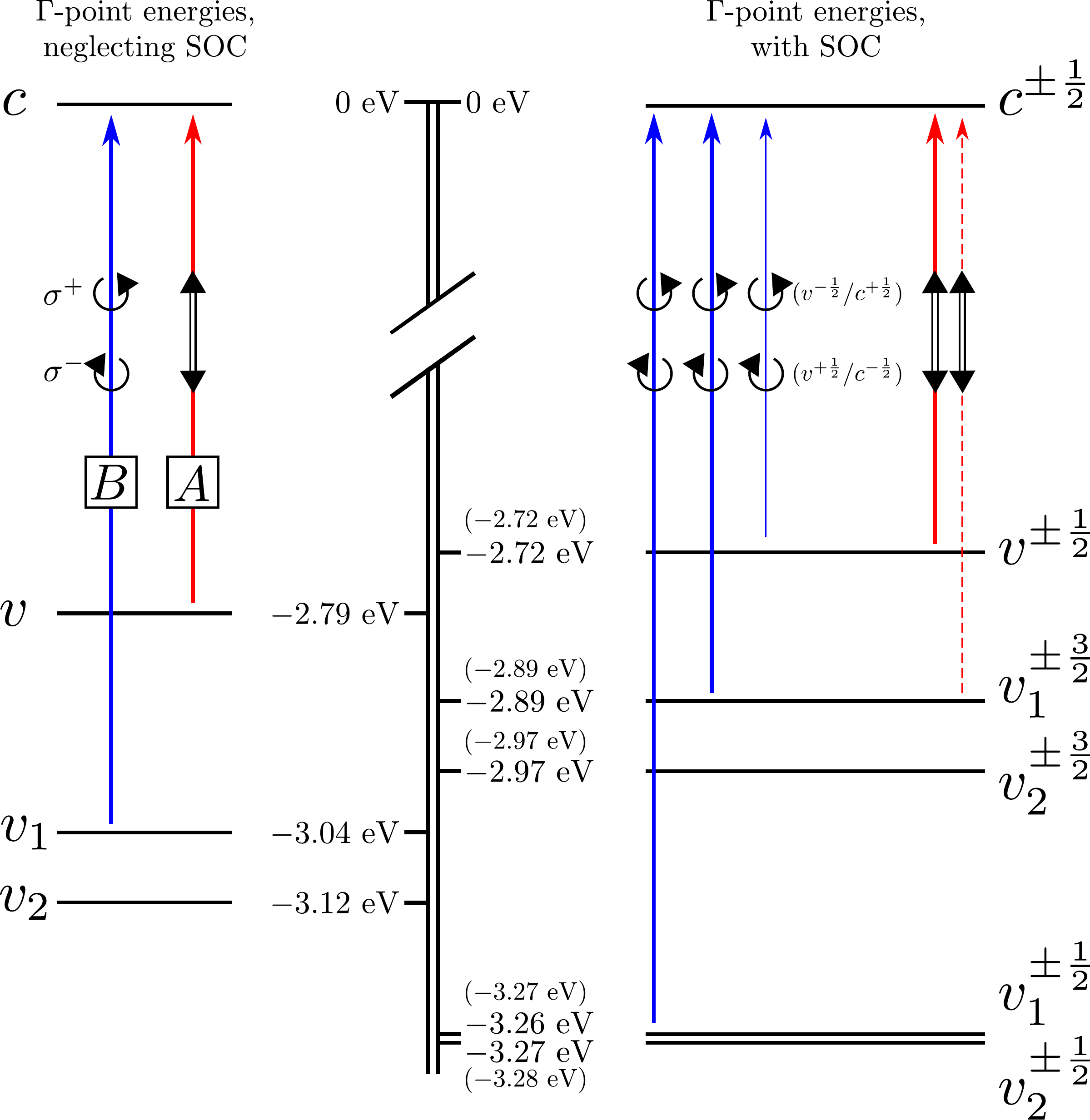}
\caption{Left hand side - Schematic of low-energy bands at $\Gamma$ in monolayer InSe without SOC, with allowed optical transitions marked. Solid arrows denote coupling of optical transitions to in-plane polarized light, dashed arrows coupling to the out-of-plane dipole transition. Right hand side - Bands and allowed optical transitions in the presence of SOC. Superscripts denote the $z$ components of total angular momentum ($J_z$) of the bands. Energies in parentheses are scissor-corrected DFT values, for comparison with the bands given by the model Hamiltonian, Eq. \ref{eq_h0_6x6}. The conduction band edge is used as a reference energy, set at 0~eV.}
\label{fig_bands} 
\end{figure}

\begin{table}
	\begin{center}
		\caption{\label{tab_d3h}Character table for irreducible representations of point group $D_3h$ of monolayer InSe, together with $\Gamma$-point classification of bands included in basis of the $\mathbf{k\cdot p}$ Hamiltonian, Eq. (\ref{eq_h0_6x6}). }
		\begin{tabular}{c|cccccc|c}
			\hline\hline 
			$D_{3h}$ & $E$ & $2C_3$ & $3C_2^{\prime}$ & $\sigma_h$ & $2S_3$ & $3\sigma_v$ & Band \\ 
			\hline
			$A_1^{\prime}$ & 1 & 1 & 1 & 1 & 1 & 1 & $v$ \\ 
		
			$A_2^{\prime}$ & 1 & 1 & $-1$ & 1 & 1 & $-1$ &  \\ 
		
			$E^{\prime}$ & 2 & $-1$ & 0 & 2 & $-1$ & 0 & $v_2$ \\ 
		
			$A_1^{\prime\prime}$ & 1 & 1 & $1$ & $-1$ & $-1$ & $-1$ &  \\ 
		
			$A_2^{\prime\prime}$ & 1 & 1 & $-1$ & $-1$ & $-1$ & $1$ & $c$ \\ 
		
			$E^{\prime\prime}$ & 2 & $-1$ & 0 & $-2$ & 1 & 0 & $v_1$ \\ 
		
		\end{tabular}
	\end{center}
\end{table} 
Here, we use a basis of spin up/down states, $\mu\equiv\mathbf{s}\cdot\mathbf{\hat{e}_z}=\pm\frac{1}{2}$, in the low-energy bands neglecting SOC, $\{c,v,v_1^{p_x},v_1^{p_y},v_2^{p_x},v_2^{p_y}\}$, labelled in the left hand side of Fig. \ref{fig_bands} and classified according to the irreducible representations of point group $D_{3h}$ of monolayer InSe in Table \ref{tab_d3h}. Band $c$ is the lowest energy conduction band, with its quadratic $\Gamma$-point dispersion described by an effective-mass single-band Hamiltonian, $H_c$. The highest-energy occupied band, $v$, has a maximum offest from $\Gamma$ in the monolayer\cite{Zolyomi2014,rybkovskiy2014transition,zhou2017multiband,debbichi2015two,jalilian2017electronic,wick2015electronic,hung2017thermo}, and we therefore describe the band with a 4th order polynomial, $H_v$. The next-highest energy valence bands, $v_{1}$ and $v_2$, are dominated by $p_x,p_y$ orbitals, and are twice-degenerate at $\Gamma$. We therefore represent the bands using 2-component Hamiltonians, $\mathbf{H}_{v_{1(2)}}$, written as matrices in a space of $p_x$ and $p_y$ orbitals, with $\mathbb{1}_\sigma$  an identity matrix, and $\sigma_{x,y,z}$ the Pauli matrices. The values of the $\mathbf{k\cdot p}$ parameters listed in Table \ref{tab_params} are determined \cite{magorrian2016electronic} from fitting to DFT dispersions without SOC near $\Gamma$. The dispersions of these bands coincide with the DFT-calculated $\Gamma$-point dispersion of InSe bands\cite{zhuang2013single,rybkovskiy2014transition,magorrian2016electronic,*tb_erratum,Zolyomi2014,zhou2017multiband,debbichi2015two,jalilian2017electronic,xiao2017defects,demirci2017structural,wick2015electronic,cai2017charge,hung2017thermo,Do2015,Li2015,Nissim2017,Ayadi2017,Li2017},  but with the band gap corrected by a `scissor correction' adjustment to the bands\cite{magorrian2016electronic}.  The factors $\dfrac{e\beta_{1(2)}}{cm_e}$,  are couplings of the spin-conserving $v_1\rightarrow c$ interband transition (B-line), and of the transition between bands $v$ and $v_2$ \footnote{Although in the absence of SOC this term is not relevant for the optical properties of InSe, it becomes important for the analysis of the effect of SOC on the principal interband transition}, respectively, to in-plane polarized light described by vector potential $\mathbf{A}=(A_x,A_y)$, with $\beta_{1(2)}=|\bra{c(v)}\mathbf{P}\ket{v_1(v_2)}|$ the magnitude of the interband matrix element of the momentum operator. The matrix element $E_zd_z$ accounts for electric dipole coupling of the $c\leftrightarrow v$ transition to out-of-plane polarized light, where $d_z=e\bra{c}z\ket{v}$ is the z-component of the interband dipole operator. Similar band structure properties have been found in monolayer GaSe \cite{Zlyomi2013,Cao2015,Chen2015,Feng2016,Si2016}.

In the spin space, $\mathbb{1}_s$ is an identity matrix, while $s_{x,y,z}$ are spin operators, with the `vectors',  $\boldsymbol{\hat{s}}=(s_x,s_y)$ and $\boldsymbol{\hat{s}}^{\mathbf{T}}=\left(\begin{array}{c}s_x\\s_y\\\end{array}\right)$, introduced to achieve a compact form for representing $\hat{H}$. Parameters $\lambda_{v_{1(2)}}$ are intraband SOC constants for band $v_{1(2)}$, determined by the atomic orbital compositions of the bands and atomic SOC strengths, while $\lambda_{v,v_1}$ and $\lambda_{c,v_2}$ take into account SOC-induced hybridization between $v$ and $v_1$, and between $c$ and $v_2$, respectively.

In calculating the latter parameters, we compare the result of diagonalization of $\hat{H}$ in Eq.~(\ref{eq_h0_6x6}) with density functional theory (DFT) calculations for InSe with SOC (VASP code \cite{VASP_PhysRevB.54.11169} in the local spin density approximation) and without SOC. We employed a plane-wave basis with a cutoff energy of 600 eV and the Brillouin
zone was sampled by a $24 \times 24 \times 1$ grid for monolayer InSe, followed by a `scissor correction' adjustment of the band gap, described in Ref. \onlinecite{magorrian2016electronic}.  
 
Atomic SOC splits\cite{Li2015,zhou2017multiband} the otherwise \cite{Zolyomi2014,magorrian2016electronic} degenerate bands at $\Gamma$ ($v_1$ and $v_2$) into pairs of states with projections $J_z=\pm\frac{3}{2}$ and $J_z=\pm\frac{1}{2}$ of total angular momentum, $\mathbf{J}=\mathbf{L}+\mathbf{s}$. The `spin-flip' part of atomic SOC, $L^{\pm}s^{\mp}$, hybridizes the lower branch of $v_1$ ($J_z=\pm\frac{1}{2}$) with $v$, to the measure determined by the matrix element $\bra{p_z}L^{\pm}\ket{p_{x,y}}$ of the angular momentum operator acting between the orbitals of the same atoms contributing to the $v$ and $v_1$ band states. This pushes $v$ higher in energy and reduces the band gap. It also hybridizes $v_2$ with $c$, but the associated shift appears to be much smaller, due to a larger energy separation between these two bands. There is no mixing between $c$ and $v_1$ or between $v$ and $v_2$ as the symmetry of these bands under $\sigma_h$ reflection requires $\bra{c}L^{\pm}\ket{v_1}=\bra{v}L^{\pm}\ket{v_2}=0$..

By diagonalizing $\hat{H}$ in Eq. (\ref{eq_h0_6x6}) and comparing the band energies with those found from DFT with SOC, Table \ref{tab_params}, we find that $\lambda_{v_1}\approx\lambda_{v_2}\approx0.30$~eV and $\lambda_{v,v_1}\approx0.25$~eV. The SOC-induced changes to the band energies at $\Gamma$ are illustrated in Fig. \ref{fig_bands}, showing good agreement between the DFT and $\mathbf{k\cdot p}$ bands. Note that inspection of the wavefunction decomposition of the conduction band edge with SOC shows a negligible ($<0.1\%$) contribution from $p_{x,y}$ orbitals (which are present due to the hybridization of $c$ and $v_2$), so we neglect the effect of $\lambda_{c,v_2}$ in our analysis, and use the conduction band edge as a reference energy, set at 0 eV. In contrast, $v-v_1$ hybridization is retained in the model as a significant effect. We find that perturbation theory overestimates the weight of $v_1$ band $p_x,p_y$ orbitals admixed into $v$, 
\begin{equation}
\label{eq_v1_v_pt}
|\delta C_v(v_1)|^2=\bigg|\frac{\frac{1}{\sqrt{2}}\lambda_{v,v_1}}{E_v-(E_{v_1}-\lambda_{v_1}/2)}\bigg|^2\sim 0.2,
\end{equation}
as compared to $|\delta C_v(v_1)|^2=0.13$ found from numerical exact diagonalization of the Hamiltonian in Eq. (\ref{eq_h0_6x6}).
\begin{table}[h]
	\caption{\label{tab_params}(a) Band energies at $\Gamma$ and model parameters for $\hat{H}$, Eq. (\ref{eq_h0_6x6}), found using a DFT calculation without SOC\cite{magorrian2016electronic}.\\ (b) Band energies from a DFT calculation with SOC included, superscripts denote the $z$ components of total angular momentum ($J_z$) of the bands. (c) SOC constants for $\hat{H}$, Eq. (\ref{eq_h0_6x6}), determined from fitting of $\Gamma$-point bands from diagonalized Hamiltonian to DFT energies with and without SOC.}
	\begin{center}
		\begin{tabular}{cc}
			\hline
			\hline
			\multicolumn{2}{c}{(a)}\\
			\hline
			$E_c$ & 0~eV\\
			$m_c$ & 0.19~$m_e$\\
			$E_v$ & $-2.79$~eV\\
			$E_2$ & $2.92$~eV\AA$^2$\\
			$E_4$ & $-38.06$~eV\AA$^4$\\
			$E_{v_1}$ & $-3.04$~eV\\
			$E_{v_2}$ & $-3.12$~eV\\
			$m_1$ & $-0.31$~$m_e$\\
			$m_2$ & $-0.30$~$m_e$\\
			$m_{1,2}^{\prime}$ & $-0.45$~$m_e$\\
			$\beta_1$&$1.09$~$\hbar/\mathrm{\AA}$\\
			$\beta_2$&$0.21$~$\hbar/\mathrm{\AA}$\\
			$d_z$&$1.68$~$e$\AA\\

			\hline
			\hline
		\end{tabular}
	\quad
			\begin{tabular}{cc}
		\hline
		\hline
		\multicolumn{2}{c}{(b)}\\
		\hline
		$E_c^{\pm\frac{1}{2}}$ & 0~eV\\
		$E_v^{\pm\frac{1}{2}}$ & $-2.72$~eV\\
		$E_{v_1}^{\pm\frac{3}{2}}$ & $-2.89$~eV \\
		$E_{v_1}^{\pm\frac{1}{2}}$ & $-3.27$~eV \\
		$E_{v_2}^{\pm\frac{3}{2}}$ & $-2.97$~eV\\
		$E_{v_2}^{\pm\frac{1}{2}}$ & $-3.28$~eV\\

		\hline
		\hline
	\end{tabular}
	\quad
			\begin{tabular}{cc}
		\hline
		\hline
		\multicolumn{2}{c}{(c)}\\
		\hline
		$\lambda_{v_1}$ & 0.30~eV\\
		$\lambda_{v_2}$ & 0.30~eV\\
		$\lambda_{v,v_1}$ & 0.25~eV\\

		\hline
		\hline
	\end{tabular}
	\end{center}
\end{table}

In the absence of SOC, the coupling of the principal interband transition, $v\leftrightarrow c$, to in-plane polarized light is forbidden due to the opposite symmetry of $c$ and $v$ under $\sigma_h$, and also due to the lack of an orbital angular momentum difference between the bands. The coupling of $v$ to the lower branch of $v_1$ by SOC relaxes this selection rule, allowing coupling to in-plane polarized light. The transitions will be between total angular momentum states $v^{\pm\frac{1}{2}}\leftrightarrow c^{\mp\frac{1}{2}}$, coupling to $\sigma^{\mp}$-polarized photons. Since $c$ changes negligibly on application of SOC, we can consider the spin-projection $\mu$ to remain a good quantum number in the conduction band, and make the observation that $\sigma^{+/-}$-polarized light will excite electrons into spin up/down states in the conduction band, as sketched in the right hand side of Fig. \ref{fig_bands}.

For coupling of the principal interband transition to in-plane polarized light, we can estimate it as
\begin{equation}
a_{SO}=\frac{e|\delta C_v(v_1)|\beta_1}{cm_e}\mathbf{A^{\pm}}\equiv\frac{e\beta_{\mathrm{sf}}}{cm_e}\mathbf{A^{\pm}},
\end{equation}
where $\mathbf{A^{\pm}}=A(\mathbf{\hat{x}}\pm i\mathbf{\hat{y}})/\sqrt{2}$ correspond to the vector potential of $\sigma^{\pm}$ circularly polarized light, and $|C_v(v_1)|^2$ is the weight of $p_x,p_y$ orbitals from $v_1$  admixed in the dominantly $p_z$ orbital based $v$-band wavefunction. The oscillator strength of such transitions can be estimated as  $\beta_{\mathrm{sf}}=|C_v(v_1)|\beta_1\approx0.4~\hbar/$\AA~ leading to the absorption coefficient\cite{magorrian2016electronic} for $\sigma^{\pm}$ light incident perpendicular to the 2D crystal,
\begin{equation}
g_{A\mathrm{(\sigma^{\pm})}}=8 \pi \frac{e^2}{\hbar c}|\beta_{\mathrm{sf}}|^2 \frac{m_c}{|E_v^{\pm\frac{1}{2}}|m_e^2}\approx 1.5\%.
\end{equation} For comparison \cite{magorrian2016electronic}, a photon incident at an angle $\theta\approx 45^{\circ}$ to the surface and coupled to the the principal interband transition via the interband out-of-plane electric dipole moment, $d_z$, is absorbed with $g_{A\mathrm{(E_z)}}[\theta\approx 45^{\circ}]\approx 3.7\%$, while for the B-line $g_{B(\sigma^{\pm})}\approx 10\%$.
\section{Multilayers}
Going on from the monolayer\cite{zhuang2013single,rybkovskiy2014transition,magorrian2016electronic,*tb_erratum,Zolyomi2014,zhou2017multiband,debbichi2015two,jalilian2017electronic,xiao2017defects,demirci2017structural,wick2015electronic,cai2017charge,hung2017thermo,Do2015,Li2015,Nissim2017,Ayadi2017,Li2017} to $N$-layer films, interlayer hopping between successive layers of InSe splits each band into $N$ subbands, as studied earlier using DFT and tight-binding calculations\cite{magorrian2016electronic,rybkovskiy2014transition,SciRep_2016}. At the $\Gamma$-point, $v_1$ and $v_2$ split very weakly, whereas $c$ and $v$, which are dominated by $s$ and $p_z$ orbitals on In and Se, exhibit a much stronger splitting. This moves the valence band edge (the top sub-band of $v$) further away from $v_1$, reducing the effect of SOC on the band edge states. This weakening in the effect of SOC, anticipated from a perturbation theory analysis similar to Eq.~(\ref{eq_v1_v_pt}), reduces the oscillator strengths of the spin-flip interband transitions. To describe quantitatively the spin-flip transitions in a multilayer film we employ a hopping model for few-layer InSe,
\begin{widetext}
\begin{align}
\begin{split}
\label{full_ham}
\hat{H}^{(N)}&=\sum_{n}^N\sum_{\substack{\alpha,\alpha^{\prime}\\\mu,\mu^{\prime}}}\hat{H}_{\alpha\mu,\alpha^{\prime}\mu^{\prime}}a^{\dagger}_{n\alpha\mu}a_{n\alpha^{\prime}\mu^{\prime}}+\sum_{n}^{N-1}\sum_{\alpha,\mu}\delta_{\alpha}\left(a_{n\alpha\mu}^{\dagger}a_{n\alpha\mu}+a_{(n+1)\alpha\mu}^{\dagger}a_{(n+1)\alpha\mu}\right)\\
&+\sum_{n}^{N-1}\sum_{\mu}\left[t_{c}a^{\dagger}_{(n+1)c\mu}a_{nc\mu}+t_{v}a^{\dagger}_{(n+1)v\mu}a_{nv\mu}+t_{cv}\left(a^{\dagger}_{(n+1)v\mu}a_{nc\mu}-a^{\dagger}_{(n+1)c\mu}a_{nv\mu}\right)+\mathrm{H.c.}\right].
\end{split}
\end{align}
\end{widetext}
Here, the operator $a^{(\dagger)}_{n\alpha\mu}$ annihilates (creates) an electron in band $\alpha$, spin state $\mu=\pm\frac{1}{2}$, in layer $n$ of the $N$-layer crystal. The sum over $\alpha,\alpha^{\prime}$ runs twice over the bands included in the monolayer Hamiltonian $\hat{H}$ in Eq.~(\ref{eq_h0_6x6}), and $\hat{H}_{\alpha\mu,\beta\mu^{\prime}}$ are the matrix elements of $\hat{H}$ between band $\alpha$ with spin projection $\mu$ and band $\alpha^{\prime}$ with spin projection $\mu^{\prime}$. The parameters $t_c\approx0.34$~eV, $t_v\approx-0.42$~eV, and $t_{cv}\approx0.29$~eV are interlayer hoppings, and $\delta_v\approx -0.06$~eV and $\delta_c<0.01$~eV (which we neglect as it is much smaller than the interlayer hoppings) are onsite energy shifts due to interlayer potentials; all of these were determined from fitting of the $\Gamma$-point energies obtained from Eq. (\ref{full_ham}) to the DFT sub-bands in 2-5 layer InSe \cite{magorrian2016electronic}. Diagonalization of $\hat{H}^{(N)}$ allows us to find the wavefunction and hence the $N$-dependence of the absorption coefficient for in-plane polarized light for the optical transition across the principal band gap of $N$-layer InSe;
\begin{equation}
g_{A\mathrm{(\sigma^{\pm})}}(N)=8 \pi \frac{e^2}{\hbar c}|\delta C^{N}_v(v_1)\beta_1|^2 \frac{m_c(N)}{\hbar \omega(N) m_e^2},
\end{equation}
where $m_c(N)$ and $\hbar\omega(N)$ are the $N$-layer conduction-band effective mass and band gap, respectively, and $|\delta C^{N}_v(v_1)|^2$ is the total weight of all $p_x,p_y$ orbitals from sub-bands of $v_1$ admixed by SOC into the highest energy valence sub-band. Overall we find 
\begin{equation}
g_{A\mathrm{(\sigma^{\pm})}}(N=3-5)\sim 0.3\%,
\end{equation}
see Table \ref{tab_absorb}. This compares with an increase in $g_{A\mathrm{(E_z)}}(N)[\theta\approx 45^{\circ}]$ from $3.7\%$ in the monolayer to $\sim 15\%$ for large $N$, and with a roughly constant 
$g_{B(\sigma^{\pm})}(N)\sim 10\%$ absorption for the B-line \cite{magorrian2016electronic}.
\begin{table}
\caption{Band gaps ($\hbar\omega$) and absorption coefficients as a function of number of layers, $N$, for coupling of A-line transition to in-plane polarized light after application of SOC ($g_{A\mathrm{(\sigma^{\pm})}}$), in comparison with absorption coefficients for the B-line ($g_{B(\sigma^{\pm})}$) and for coupling to the out-of-plane dipole transition for a photon incident at an angle $\theta\approx 45^{\circ}$ to the crystal ($g_{A\mathrm{(E_z)}}$), taken from Ref. \onlinecite{magorrian2016electronic}. \label{tab_absorb}}
\begin{center}
\begin{tabular}{ccccc}
\hline
\hline
$N$ & $\hbar\omega$(eV)& $g_{A\mathrm{(\sigma^{\pm})}}$(\%) & $g_{A\mathrm{(E_z)}}[\theta\approx 45^{\circ}]$(\%)& $g_{B(\sigma^{\pm})}$(\%)\\
\hline
$1$ & 2.72 & 1.5 &3.7&  $10.3$\\
$2$ & 2.00 &$0.5$&6.4  & $8.3$  \\
$3$ & 1.67 &$0.3$&8.2 &   $8.8$  \\
$4$ & 1.50 &$0.3$&9.6 &   $9.0$  \\
$5$ & 1.40 &$0.3$&10.7  &  $9.1$  \\
\hline
\hline
\end{tabular}
\end{center}
\end{table}
\begin{figure}
\includegraphics[width=0.40\textwidth]{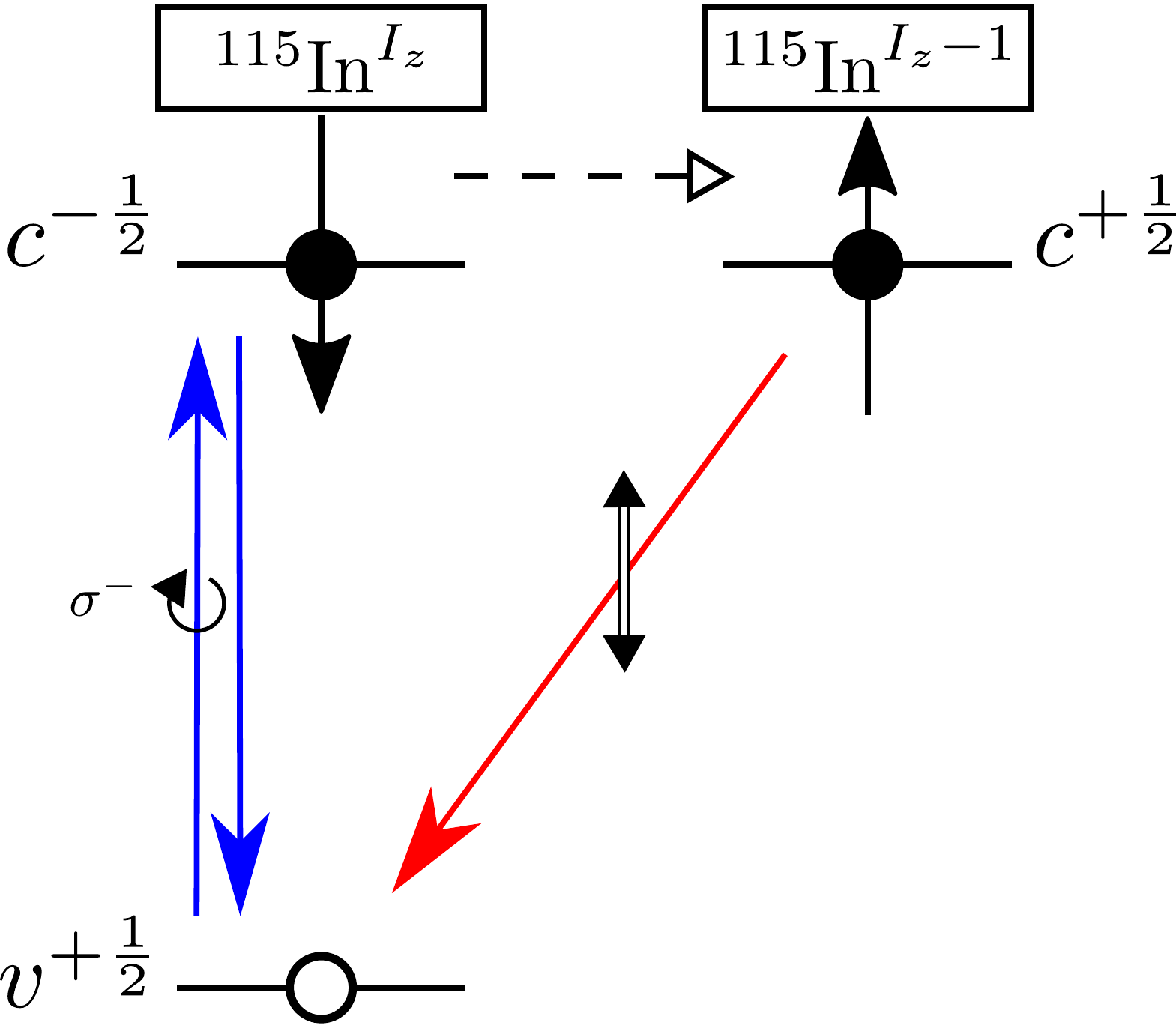}
\caption{Schematic of spin-pumping into In nuclear spin polarization - an electron excited by $\sigma^-$-polarized light goes into a spin-down state in the conduction band, and can recombine by emission of another $\sigma^-$ photon. After the electron flips its spin via hyperfine interaction with In nuclei, it can then recombine with the photoexcited hole by emitting an out-of-plane polarized photon, leaving the angular momentum imparted in the nuclear spin of the In atoms.}
\label{fig_hfs} 
\end{figure}
\section{Nuclear spin pumping}
The excitation of electrons into $\mu=\pm\frac{1}{2}$ states of the conduction band by $\sigma^{\pm}$-polarized light tuned to the energy of the principal interband transition allows for the optical pumping of In nuclear spins. This would occur through the mechanism sketched in Fig. \ref{fig_hfs}. An electron excited by $\sigma^{-}$-polarized light into the $\mu=-\frac{1}{2}$ conduction band state, leaving behind a hole in the $J_z=+\frac{1}{2}$ valence band, $v^{+\frac{1}{2}}$, can transfer its spin to the nucleus via the hyperfine interaction, $e^{\downarrow}+\mathrm{In^{I_z}}\rightarrow e^{\uparrow}+\mathrm{In^{I_z-1}}$, where $I_z$ is the $z$-component of the indium's nuclear spin. After that, it can recombine without spin-flip with the photoexcited hole, with a similar cycle to be repeated following the absorption of the next $\sigma^{-}$-polarized photon. 

A strong contribution of an indium $s$ orbital  to the conduction band states presents the possibility of a strong coupling between the electronic spin and In nuclear spins. The most common isotope of indium, $^{115}\mathrm{In}$, has a nuclear spin $I=\frac{9}{2}$, and an atomic hyperfine coupling constant $A_{\mathrm{In}}\approx 60~\mu\mathrm{eV}$ \cite{In_hfs} (in comparison with Se, for which the only stable nucleus with non-zero spin, $^{77}\mathrm{Se}$, $I=\frac{1}{2}$, abundance 7\%, has a much smaller coupling constant $A_{\mathrm{Se}}\approx 2~\mu\mathrm{eV}$ \cite{Se_hfs}). Using the orbital decomposition reported in Ref.~\onlinecite{magorrian2016electronic} we can estimate an effective hyperfine coupling constant for the conduction band, $A^{eff}_{c}=|C_c(\mathrm{In}_s)|^2A\approx 15~\mu\mathrm{eV}$, where $|C_c(\mathrm{In}_s)|^2$ is the weight of In $s$ orbitals in the conduction band state near the $\Gamma$-point. As a result, one can expect that monolayers and bilayers of InSe will offer a suitable material for achieving a high degree of optically induced nuclear spin polarization. 
\section{Conclusions}
In conclusion, we have used a $\mathbf{k\cdot p}$ model to show how spin-orbit coupling in mono- and few-layer InSe allows coupling between the principal interband transition and in-plane polarized light, accompanied by a spin-flip, with the coupling at its strongest in the monolayer, $g_{A(\sigma^{\pm})}\sim 1.5\%$, saturating to $g_{A(\sigma^{\pm})}\sim 0.3\%$ for $3-5$ layers. We would expect a similar effect in the other III-VI semiconductors, such as GaSe, with the strength primarily determined by the SOC strength in the group VI atoms. Electrons, excited selectively into spin up/down states in the conduction band using $\sigma^{\pm}$-polarized light, can transfer their spin angular momentum to the In nuclei , allowing for optically induced nuclear spin polarization. The $\mathbf{k\cdot p}$ model presented here can be amended to include the effect on optical transitions in few-layer InSe of an applied displacement field in a dielectric environment, through changes to the onsite band energies and coupling to states in the dielectric environment, and this will be addressed in future work.
\begin{acknowledgments}
	The authors thank A. Patan\`e, A. V. Tyurnina,  M. Potemski, Y. Ye, J. Lischner, and N. D. Drummond for discussions. This work made use of the facilities the CSF cluster of the University of Manchester. SJM acknowledges support from EPSRC CDT Graphene NOWNANO EP/L01548X. VF acknowledges support from ERC Synergy Grant Hetero2D, EPSRC EP/N010345, and Lloyd Register Foundation Nanotechnology grant. VZ and VF acknowledge support from the European Graphene Flagship Project, the N8 Polaris service, the use of the ARCHER national UK supercomputer (RAP Project e547), and the Tianhe-2 Supercomputer at NUDT. Research data is available from the authors on request.
	
\end{acknowledgments}
%

\end{document}